\documentclass[aps,prl,floatfix,twocolumn,showpacs,amsmath,amssymb]{revtex4}
\usepackage{CJK}
\usepackage{graphicx}
\usepackage{epsfig}
\usepackage{latexsym}
\usepackage{amsmath}
\usepackage{wasysym}
\usepackage{nicefrac}
\usepackage{color}
\usepackage{dcolumn}
\usepackage{bm}
\usepackage{footmisc}
\begin{document}

\title{Phase diagram of the triangular extended Hubbard model}
\author{Luca F. Tocchio\footnote{Present address: SISSA, via Bonomea~265, 34136~Trieste, Italy}}
\affiliation{Institute for Theoretical Physics, University of Frankfurt, 
Max-von-Laue-Stra{\ss}e 1, D-60438 Frankfurt a.M., Germany} 
\author{Claudius Gros}
\affiliation{Institute for Theoretical Physics, University of Frankfurt, 
Max-von-Laue-Stra{\ss}e 1, D-60438 Frankfurt a.M., Germany} 
\author{Xue-Feng Zhang}
\affiliation{Physics Department and Research Center OPTIMAS,
University of Kaiserslautern, D-67663 Kaiserslautern, Germany}
\author{Sebastian Eggert}
\affiliation{Physics Department and Research Center OPTIMAS,
University of Kaiserslautern, D-67663 Kaiserslautern, Germany}

\date{\today} 

\begin{abstract}
We study the extended Hubbard model on the triangular lattice 
as a function of filling and interaction strength. The complex 
interplay of kinetic frustration and strong interactions on the 
triangular lattice leads to exotic phases where long-range charge 
order, antiferromagnetic order, and metallic conductivity can 
coexist. Variational Monte Carlo simulations show that three 
kinds of ordered metallic states are stable as a function of 
nearest neighbor interaction and filling. The coexistence of 
conductivity and order is explained by a separation into two 
functional classes of particles: part of them contributes to 
the stable order, while the other part forms a partially filled 
band on the remaining substructure. The relation to charge 
ordering in charge transfer salts is discussed.
\end{abstract}


\maketitle


The study of frustrated and strongly interacting systems in two dimensions
has received an unbroken intensity of research activities in recent years.
Just to name a few examples, spin-liquid states have been postulated in 
frustrated two-dimensional antiferromagnets \cite{balents10,yan11,iqbal13,tocchio13}, 
supersolid phases have been established for hard-core bosons on a triangular 
lattice \cite{wessel05,melko05,heidarian05}, and the concept of deconfined 
quantum critical points \cite{senthil04} has sparked a tremendous interest 
in the search of exotic phase transitions.

\begin{figure}
\includegraphics[width=0.8\columnwidth]{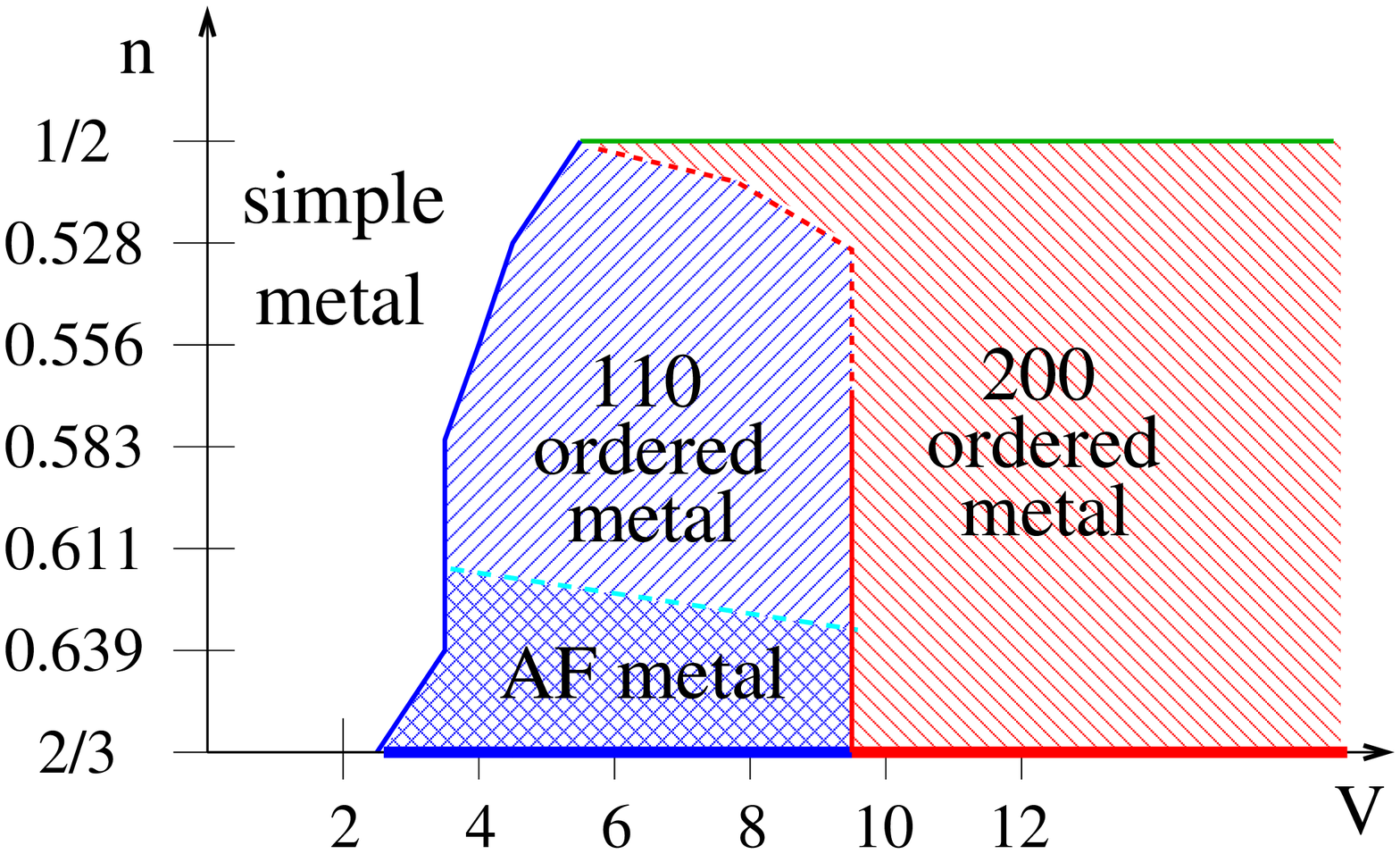}

\vspace{0.2cm}
\includegraphics[width=0.7\columnwidth]{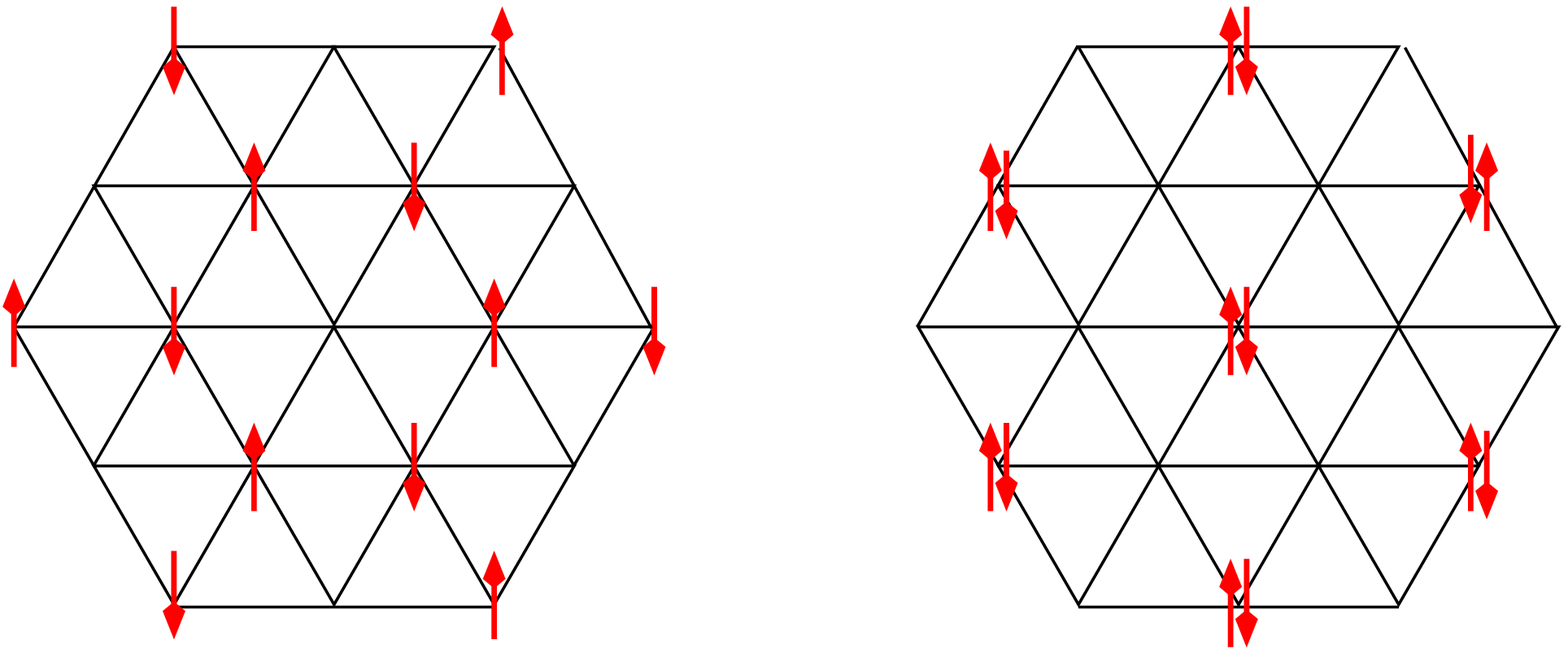}
\caption{\label{fig:pd} (Color online) 
Upper panel: Phase diagram of the Hubbard model for $U=30t$ as a 
function of $V/t$ and filling. We indentify 
a simple metallic phase, a metallic state with a 110 charge order, and 
a metallic state with a 200 charge order. The charge order 
regions are insulating at commensurate filling $n=2/3$ 
(thick lines) and metallic otherwise. In the limiting case 
$n=1/2$ (green line) only the 200 charge ordered phase 
appears, with increasing electron occupation on one 
sublattice as a function of $V/t$. For filling 
$0.62\lesssim n\le 2/3$ we find an indication for magnetic 
order (AF) within the 110 charge ordered metal.
Lower panel: sketch of the 110 charge order (left) and of 
the 200 charge order (right).}
\end{figure}

Above examples involve essentially spin-like systems, where charge degrees of 
freedom only play a passive role. When both spin and charge degrees of 
freedom are considered at incommensurate filling the situation 
potentially becomes even more interesting. In order to study the interplay of
frustration and strong interactions with spin and charge degrees of freedom
at arbitrary filling $n= \langle n_\uparrow + n_\downarrow \rangle$,
the extended Hubbard model on the triangular lattice is the prototypical system to study. 
In standard notation the model is given by the many body Hamiltonian 
\begin{equation}\label{eq:hubbard}
{\cal H}=-t\sum_{\langle i,j\rangle,\sigma} c^\dagger_{i,\sigma} c_{j,\sigma} + 
\textrm{h.c.} + U \sum_{i} n_{i,\uparrow} n_{i,\downarrow}+V \sum_{\langle i,j\rangle} n_i n_j.
\end{equation}
In addition to its fundamental importance, 
this model is of interest for describing the rich and complex behavior of 
organic conductors, such as the charge transfer salts 
$\theta$-(BEDT-TTF)$_2$X 
\cite{hotta03,udagawa07,canocortes11,merino13,mori03,watanabe06,watanabe05,merino07}, 
where molecules are arranged on an anisotropic triangular lattice with 
incommensurate filling.  The Hubbard model on the triangular lattice 
has also been considered in the context of explaining superconductivity 
in the layered compound Na$_x$CoO$_2$ \cite{takada03,honerkamp03} 
where interesting textures have been predicted recently~\cite{jiang13}. 
Unfortunately, analytical and numerical studies of this model are far 
from trivial and to our knowledge it has not yet been analyzed with 
quantum many body simulations for incommensurate filling. In this paper 
we now use numerical variational Monte Carlo simulations in order to 
establish the phase diagram as a function of filling and interaction 
strength. In addition to the ordinary metallic phase, three interesting 
phases are found, where long-range charge order and metallic conductivity 
are present simultaneously as depicted in Fig.~\ref{fig:pd}, which summarizes 
most of our findings. 

In order to systematically understand the different phases let us 
first review the simpler case of commensurate 2/3 filling 
($n=2/3$) discussed in Ref.~\cite{watanabe05}. Quite intuitively, 
for strong $V\agt U/3 \gg t$ any nearest neighbor occupation is 
forbidden, resulting in an insulating ordered phase with exactly 
two electrons on one sublattice (200 order), while for weaker 
nearest neighbor repulsion double occupancy is forbidden and 
instead two sublattices are half-filled in an hexagonal order 
(110 order), see Fig.~\ref{fig:pd}. Neglecting hopping, these 
phases are also stable for lower filling \cite{mori03}. However, 
using a grand-canonical point of view, the filling simply jumps 
as a function of chemical potential $\mu$ in the limit of vanishing 
hopping. In particular, at $\mu=3V$ a jump occurs from the stable 
100-order with 1/3 filling to a 2/3 filled 110-state. At finite 
hopping $t\ll V\ll U$, however, incommensurate filling is stabilized 
in a finite range $|\mu -3V|\alt 3t$ due to the kinetic energy of 
particles (holes) on the hexagonal order \cite{zhang11}. 
In the following we will focus on analyzing the partially
filled state in 
the density range $1/2 \leq n \leq 2/3$. 

For incommensurate filling, the 110 phase in Fig.~\ref{fig:pd} 
represents a state where one of the three triangular sublattices 
remains empty and all the electrons occupy the other two 
sublattices in an hexagonal density order in order to minimize 
the nearest neighbor repulsion. Since the hexagonal order 
necessarily contains holes for $n < 2/3$, this phase becomes
conducting.
Interestingly, this coexistence of two counter-intuitive properties 
(order and conductivity) is directly related to the supersolid
state on triangular lattices which has been established for
hardcore bosons \cite{wessel05,heidarian05,melko05} and has
analogously been postulated for spinless fermions \cite{hotta06, hotta07}.
Simulations for hardcore bosons have shown that a separation
into two types of hole-like particles is possible \cite{zhang11}: 
One part creates the ordered state by keeping one sublattice empty, 
while the other part can move freely on the hexagonal structure (partial liquid). 
However, as the filling approaches $n=2/3$ we observe a transition 
to antiferromagnetic (AF) order coexisting with conducting behavior, which must have a 
different mechanism as described below.

For the extended Hubbard model we find a third interesting phase 
in the form of a 200 order in Fig.~\ref{fig:pd}:
Double occupancy occurs only on one sublattice which 
is reduced with filling and gives an ordered state.
This state has some surprising properties, 
since the observed conductivity implies that 
the other two sublattices are not completely empty either.
The occupation on those two sublattices therefore {\it increases} 
with decreasing filling, which leads to conductive behavior. 
The phase transition between the 110 and 200 states is first order close to 
commensurate filling but may become second order for $n\alt 0.57$.
For large hoppings a transition to a simple metal occurs.

\begin{figure}
\includegraphics[width=0.8\columnwidth]{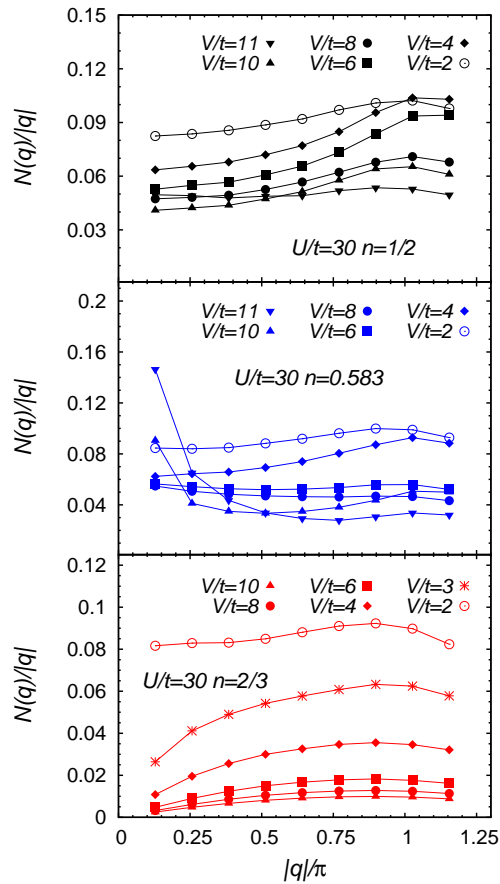}
\caption{\label{fig:MIT} (Color online) Lower panel: $N(q)/q$ as a function of $|q|/\pi$ for different values of $V/t$ 
and doping $n=2/3$. Data are shown along the path in the Brillouin zone connecting the point $Q=(0,2\pi/\sqrt{3})$ 
to the point $\Gamma=(0,0)$. The on-site Coulomb repulsion is $U=30t$ and data refer to the $L=342$ case. Middle panel: 
The same as the lower panel but with $n=0.583$. Upper panel: The same as the lower panel but with $n=1/2$.}
\end{figure}

In order to simulate the model in Eq.~(\ref{eq:hubbard}) at 
zero temperature we have used the Variational 
Monte Carlo (VMC) method~\cite{vmc}. This method gives 
very good results~\cite{imada2013} even for correlated and 
frustrated systems by numerically sampling expectation 
values over a variational ansatz. A powerful correlated variational 
state is given by \cite{grosbcs,zhang,grosAnnals,capello_Jastrow}
$|\Psi_{\textrm{FS}}\rangle = {\cal J}
|\textrm{FS}\rangle$, where $|\rm{FS}\rangle$ is the non-interacting
filled Fermi sea, to which a finite small superconductive term is added 
in order to regularize the wave function, i.e.\ to separate the 
highest occupied and the lowest unoccupied 
states by a gap. The term ${\cal J}=\exp(-1/2 \sum_{ij} v_{ij} n_i
n_j)$ is a density-density Jastrow factor, where the $v_{ij}$'s 
are optimized with VMC for
every independent distance $|i-j|$ (including on-site). 
Backflow correlations further improve the correlated
state $|\Psi_{\rm{FS}}\rangle$;
in this approach, each orbital that
defines the unprojected state $|\textrm{FS}\rangle$ is taken to
depend upon the many-body configuration in order to incorporate
virtual hopping processes~\cite{backflow}.
The non-interacting state $|\rm{FS}\rangle$ also includes
three different chemical potentials as variational parameter, one 
for each sublattice. We must emphasize however, that even for a 
uniform variational chemical potential the charge ordered metallic 
states spontaneously appears in the phase diagram at arbitrary filling, 
which demonstrates the stability of this phenomenon. Finally, a 
coupling to an external field can be added to the mean-field 
$|\textrm{FS}\rangle$ state in order to check if the ground 
state is magnetically ordered. All results presented here
are obtained by fully incorporating the backflow corrections and
optimizing individually~\cite{sorella} every variational parameter 
in the wave function. 

The static structure factor $N(q)=\langle n_{-q}n_q\rangle$ is 
a good indicator for metallic behavior, where 
$n_q=1/\sqrt{L}\sum_{r,\sigma} e^{iqr}n_{r,\sigma}$ is the Fourier 
transform of the particle density. The metallic phase is characterized 
by $N(q)\propto q$ for $q\to 0$, which implies a vanishing gap for 
particle-hole excitations. On the contrary, $N(q)\propto q^2$ for 
$q \to 0$, implies a finite charge gap and insulating 
behavior.~\cite{backflow} We find conducting behavior everywhere 
except for $n=2/3$ and $V/t \gtrsim 3$, as shown in Fig.~\ref{fig:MIT}.
Interestingly, a diverging behavior of $N(q\to0)$ is observed in the $200$ phase,
which we attribute to a $q^2$ dispersion relation at effective low filling
as explained below. 
\begin{figure}
\includegraphics[width=0.8\columnwidth]{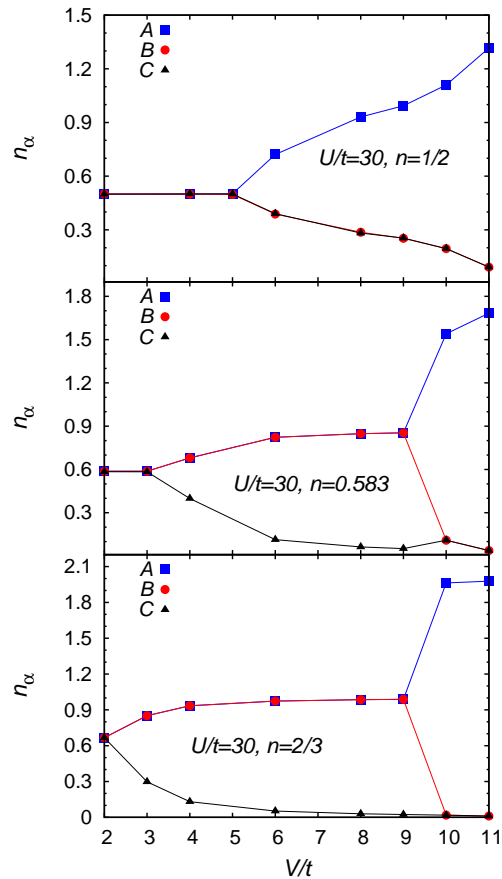}
\caption{\label{fig:density} (Color online) Electronic density $n_{\alpha}$ in each of the three 
sublattices $A$, $B$, and $C$ as a function of $V/t$. Data are presented for the three values of the total electronic density 
$n=1/2$, $n=0.583$, and $n=2/3$. The on-site Coulomb repulsion is $U=30t$ and the lattice size is $L=324$.}
\end{figure}

In order to distinguish between the different kinds of charge 
ordering in the model, we plot in Fig.~\ref{fig:density} the 
electronic density per sublattice $n_{\alpha}$ with $\alpha=A,B,C$, that is 
the number of electrons divided by the number of sites, in each 
of the three sublattices. Within the non-ordered metallic phase, 
the electronic density is expected to be the same on each sublattice, 
while in the 110 region one sublattice depletes, with the electrons 
forming an hexagonal density order, see Fig.~\ref{fig:pd}. Finally, 
in the 200 phase one sublattice is occupied with a density much greater 
than one, due to the large number of double occupancies. Both the cases 
$n=2/3$ and $n=0.583$ in Fig.~\ref{fig:density} show a clear distinction 
between the three regimes, while in the limiting case $n=1/2$ only the 
200 charge order may be observed, with a single sublattice being more 
and more occupied as long as the ratio $V/t$ increases. 

\begin{figure}
\includegraphics[width=0.8\columnwidth]{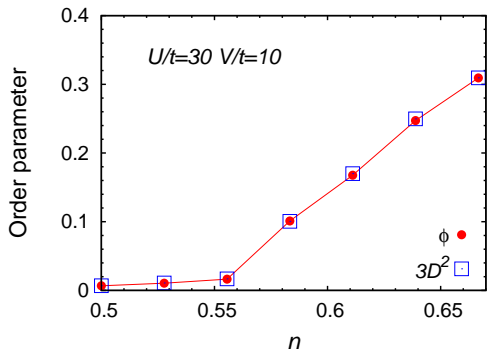}
\caption{\label{fig:phi} (Color online) Charge order parameter $\phi$, see Eq.~(\ref{eq:phi}), 
and $3D^2$, where $D$ is the density of double occupancies, 
as a function of the electronic density $n$. Data are presented at $U=30t$, $V=10t$ and for a lattice size $L=324$.}
\end{figure}

\begin{figure}
\includegraphics[width=0.8\columnwidth]{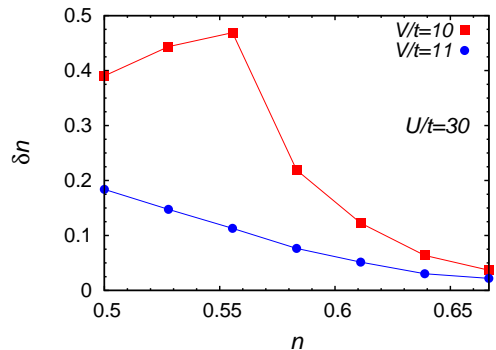}
\caption{\label{fig:deltan} (Color online) Filling of the hexagonal substructure $\delta n$ 
as a function of the total electronic density $n$ within the 200 region of the phase diagram, at $V/t=10$ and $V/t=11$. 
Data are shown at $U=30t$ on a $L=324$ lattice size.}
\end{figure}

Even though an high density of electrons on one sublattice 
is the expected behavior for a small hopping $t$ and $3V>U$, 
the 200 ordered metal has rather unusual properties. First 
of all it is far from obvious why this ordered state is 
conducting. In the 110 order the conductivity can be 
explained by mobile holes moving on a hexagonal substructure 
\cite{hotta06,hotta07,zhang11}.
In the 200 order on the other hand, holes only appear on one 
sublattice which is not connected, so this argument fails.
Moreover, we have checked that all the electrons that are 
in a doubly occupied state also contribute to the ordering, 
so conduction by virtual hopping or by pair hopping can be 
ruled out. We present in Fig.~\ref{fig:phi} the charge order 
parameter $\phi$ as defined by
\begin{equation}\label{eq:phi}
\phi=\lim_{|i-j|\to \infty}\langle 
(n_{i,\uparrow}n_{i,\downarrow})(n_{j,\uparrow}n_{j,\downarrow})\rangle\,,
\end{equation}
where the distance $|i-j|$ connects points on the same sublattice, 
and compare it with the density of double occupancies 
$D=\langle n_{\uparrow}n_{\downarrow}\rangle$. 
If the relation $\phi=3D^2$ holds, all the double occupancies 
participate to the charge order, otherwise, if $\phi<3D^2$, 
a fraction of the double occupancies is mobile outside the
200 pattern, with the limiting case $\phi=D^2$ corresponding 
to an uniform distribution of double occupancies in the lattice. 
According to the result 
shown in Fig.~\ref{fig:phi} for $V/t=10$, the relation 
$\phi=3D^2$ is verified in all the doping range 
and the system separates into charge ordered double 
occupancies and free electrons that are responsible 
for the conduction mechanism. Therefore, conductivity 
appears to require a small density of electrons on the 
two sublattices which are empty for $n=2/3$, i.e.\ the 
density on the two almost empty sublattices must
{\it increase} with decreasing $n$. In Fig.~\ref{fig:deltan} 
we show $\delta n$, that represents the electronic 
filling on the hexagonal substructure, which is 
available for a conducting band in the 200 regime. 
In the case $V/t=11$ it is clear that $\delta n$ 
increases at increasing doping, while in the case 
$V/t=10$ there is a small decrease in $\delta n$ when 
doping becomes large, i.e. $n \lesssim 0.57$. This is 
just a consequence of the small and almost constant 
number of double occupancies that occurs at $V/t=10$ 
in the range $1/2\le n \lesssim 0.57$, see Fig.~\ref{fig:phi}. 
Indeed, in this density range the transition between 
the 110 and the 200 phases becomes second order with a 
smooth increase of the number of double occupancies 
as a function of $V/t$. The effective filling of electrons 
on the hexagonal substructure $\delta n$ is rather low 
in the range $0.57\lesssim n \le 2/3$. Accordingly, the electrons
follow a $q^2$ dispersion relation 
at the bottom of the band, which explains the
divergence of $N(q\to0)$ in this phase, as discussed above in Fig.~\ref{fig:MIT}.

Finally, we also tested for magnetic order and found that an antiferromagnetic state 
has lower variational energy for fillings $0.62\lesssim n\le 2/3$ 
as indicated in Fig.~\ref{fig:pd}.
While antiferromagnetic order is expected for commensurate insulating fillings, it should
immediately be destroyed by moving holes on the hexagonal substructure.
However, for very small doping close to filling $2/3$ the energy gain from 
hopping of order $t(2/3-n)$ is not sufficient to overcome the energy gain from long-range
antiferromagnetic order of order $n t^2/U$. Nonetheless, {\it second order} hopping 
processes of holes via the depleted sites 
are still possible without changing the spin orientation, so that a finite
conductivity is observed in co-existence with antiferromagnetic order in this special 
case. This phase is stabilized for larger second order hopping amplitude $t^2/V$
in agreement with our finding in Fig.~\ref{fig:pd}.

In conclusion, we have analyzed the extended Hubbard model on 
the triangular lattice as a function of interaction strength 
and filling. The phase diagram in Fig.~\ref{fig:pd} shows three 
metallic phases at incommensurate filling. A simple metallic phase 
is confirmed for large hopping. With increasing interaction strength 
an ordered metal with a 110-type order is observed, due to the 
appearance of holes on a stable hexagonal order, which is analogous 
to the underlying mechanism for supersolidity \cite{zhang11}. For 
filling close to $2/3$ we observe a phase transition to an 
antiferromagnetically ordered metal. A 200-ordered phase with 
one double occupied sublattice is found for still larger nearest 
neighbor repulsion, which surprisingly also shows conductive behavior.
The observed occupancy of the sublattices B and C and the electronic 
properties are consistent with a band on the hexagonal substructure 
with very low filling. This is surprising, since 
the strong nearest neighbor repulsion naively presents a large energy barrier for
electrons on the hexagonal substructure next to the double occupied sites. 
The detailed mechanisms of the conductive behavior both in the 200 phase and
in the 110 antiferromagnetic phase remain a topic of future research.

Experimentally, charge ordering phenomena in charge transfer salts 
have been researched with a large variety of methods, e.g. NMR, 
X-ray and Infrared/Raman spectroscopy \cite{takahashi06}. Coexistence 
of metallic behavior and charge ordering has only been observed in 
few cases for $\theta$-(BEDT-TTF)$_2$X and $\beta''$-(BEDT-TTF)(TCNQ) 
charge transfer salts and only for short range charge order \cite{yakushi12}.
The scenario we have proposed in this paper predicts a coexistence 
of metallic behavior and {\it long-range} order, which is not due 
to a partial instability of the Fermi surface. Instead we can identify 
a separation into two functional classes of particles (or holes): part 
of them contribute to a stable order on one sublattice, while another 
part forms a partially filled band on the remaining hexagonal 
substructure.

We would like to thank the Deutsche Forschungsgemeinschaft for
financial support through grant SFB/TR49 and Federico Becca 
for useful discussions.


%
%
\end{document}